\documentclass[english,manuscript,superscriptaddress]{revtex4}
\usepackage[T1]{fontenc}
\usepackage[latin9]{inputenc}
\usepackage{color}
\usepackage{textcomp}
\usepackage{amsmath}
\usepackage{amssymb}

\makeatletter

\DeclareRobustCommand{\greektext}{%
  \fontencoding{LGR}\selectfont\def\encodingdefault{LGR}}
\DeclareRobustCommand{\textgreek}[1]{\leavevmode{\greektext #1}}
\DeclareFontEncoding{LGR}{}{}
\DeclareTextSymbol{\~}{LGR}{126}
\newcommand{\lyxmathsym}[1]{\ifmmode\begingroup\def\b@ld{bold}
  \text{\ifx\math@version\b@ld\bfseries\fi#1}\endgroup\else#1\fi}

\providecommand{\tabularnewline}{\\}

\@ifundefined{textcolor}{}
{%
 \definecolor{BLACK}{gray}{0}
 \definecolor{WHITE}{gray}{1}
 \definecolor{RED}{rgb}{1,0,0}
 \definecolor{GREEN}{rgb}{0,1,0}
 \definecolor{BLUE}{rgb}{0,0,1}
 \definecolor{CYAN}{cmyk}{1,0,0,0}
 \definecolor{MAGENTA}{cmyk}{0,1,0,0}
 \definecolor{YELLOW}{cmyk}{0,0,1,0}
 }

\makeatother

\usepackage{babel}
\begin{document}

\title{\textcolor{black}{Compensation of B-L charge of matter with relic
sneutrinos}}

\author{\textcolor{black}{S.O. Kara}}

\email{sokara@science.ankara.edu.tr}

\affiliation{\textcolor{black}{Ankara University, Physics Department, Ankara,
Turkey}}

\author{\textcolor{black}{S. Turkoz}}

\email{turkoz@science.ankara.edu.tr}

\affiliation{\textcolor{black}{Ankara University, Physics Department, Ankara,
Turkey}}
\begin{abstract}
\textcolor{black}{We consider massless gauge boson connected to B-L
charge with and without compensation to complete the investigation
of the gauging of B and L charges. Relic sneutrinos predicted by SUSY
and composite models may compensate B-L charge of matter. As a consequence
of this possible compensation mechanism we have shown that the available
experimental data admit the range of the B-L interaction constant,
${\color{blue}{\color{black}10^{-29}<\alpha_{B-L}<10^{-12},}}$ in
addition to }\textcolor{black}{\normalsize $\alpha_{B-L}<10^{-49}$
obtained without compensation.}\textcolor{black}{{} }
\end{abstract}
\maketitle
\textcolor{black}{Recently, there has been renewed interest in gauging
baryon (B) and lepton (L) numbers \cite{Perez,Dulaney,Perez2,Ko,Perez3,Kara}.
Historically these numbers have been proposed to explain non observation
of certain processes such as proton and neutrinoless double beta decays.
By analogy with QED it is natural to consider possible existence of
gauge bosons connected to baryon and lepton charges. Massless baryonic
and leptonic photons has been considered long time ago by T.D. Lee
and C.N. Yang \cite{Lee} and L.B. Okun \cite{Okun}, respectively.
However, the experiments on checking the equality of inertial and
gravitational masses by Eötvös and his colleagues \cite{Eotvos} have
put very strong upper limits on the value of baryonic and leptonic
coupling constants: $\alpha_{B}<10^{-47}$, $\alpha_{L}<10^{-49}$
\cite{Okun2}. By comparing these values with electromagnetic coupling
constant, $\alpha_{em}\sim10^{-2}$, it has been concluded that the
existence of such baryonic and leptonic photons are unrealistic. These
statements can be weakened, if the mechanism for compensation of the
baryonic and leptonic charges of matter exists. In \cite{Ciftci}
has been shown that leptonic charge of matter can be compensated by
the relic scalar anti-neutrinos predicted by SUSY or composite (preonic)
models \cite{Weinberg,Goldberg,Krauss,Ellis,D'Souza,Buchmann,Harari,Shupe}.}

\textcolor{black}{Concerning the baryon charge alone there is no realistic
candidate which can provide compensation. Massive gauge bosons connected
to baryon and lepton charges have been considered in \cite{Foot,Rajpoot,Rajpoot2,He,Carone,Carone2}.
Massive gauge boson connected to B-L is considered in \cite{Basso,Basso2,Buchmuller,Khalil}.
To complete the gauging of B and L charges in this paper we consider
massless gauge boson connected to B-L charge with and without compensation.}

\textcolor{black}{In case of the massless gauge boson, the upper limits
on the equality of inertial and gravitational masses obtained by Eötvös
\cite{Eotvos}, improved by Braginsky and Panov \cite{Braginsky}
and updated by Eöt-Wash group from Washington University \cite{Schlamminger,Adelberger}
give a restriction on the strength of $\alpha_{B-L}$ coupling constant.
By adopting the expression given in ref \cite{Okun3}}

\textcolor{black}{
\begin{equation}
\frac{\alpha_{B-L}}{G}\left(\frac{A-Z}{M}\right)\left\{ \frac{\left(A-Z\right)_{Be}}{M_{Be}}-\frac{\left(A-Z\right)_{Ti}}{M_{Ti}}\right\} <{\color{black}{\color{blue}{\color{blue}{\color{red}{\color{blue}{\color{black}\{}}\begin{array}{c}
{\color{black}{\color{black}{\color{blue}{\color{black}(0.3\pm1.8)\times10^{-13}Earth}}}}\\
{\color{black}{\color{black}{\color{blue}{\color{black}(-3.1\pm4.7)\times10^{-13}Sun}}}}
\end{array}}}}}
\end{equation}
where $\alpha_{B-L}$ is interaction constant for B-L charge, A, Z,
M are baryon number, atomic number and mass respectively, of the corresponding
objects, namely earth, sun and test objects made by Beryllium and
Titanium, G=$6.10^{-39}m_{p}^{-2}$ is the Newton gravitational constant
($m_{p}$ = mass of proton) we obtain upper limiting values for B-L
coupling constant $\lyxmathsym{\textgreek{a}}_{B-L}$, namely,}

\textcolor{black}{
\begin{equation}
\alpha_{B-L}<\{{\color{red}\begin{array}{c}
{\color{black}{\color{red}{\color{black}2.1\times10^{-49}Earth}}}\\
{\color{black}5.3\times10^{-49}Sun{\color{white}n}}
\end{array}{\color{black}.}}
\end{equation}
In the calculation of baryon number the composition of earth and sun
has been taken into account \cite{Brown,Sun}.}

\textcolor{black}{These upper values for $\alpha_{B-L}$ indicate
that B-L photons ${\color{black}\gamma}_{{\color{black}B-L}}$ are
unrealistic. To overcome these limiting values for $\alpha_{B-L}$
we propose the compensation of the B-L charge of matter with relic
scalar neutrinos which has been regarded as dark matter candidate
\cite{Falk,Ibanez,Hagelin,Hall,Han,Hooper}. The B-L charge of matter
is solely that of neutrons since the B-L charge of proton and electron
cancels each other\textquoteright{}s effect. }

\textcolor{black}{By the analogy with electrodynamics we write down
an equation which gives the B-L potential of an external B-L charge
in relic scalar neutrino and antineutrino background \cite{Feynman}:}

\textcolor{black}{
\begin{equation}
\nabla^{2}\phi\left(x\right)=-4\pi\alpha_{B-L}\left(n_{n}\left(x\right)+n_{0}e^{-\phi\left(x\right)/kT_{0}}-n_{0}e^{\phi\left(x\right)/kT_{0}}\right)dV.
\end{equation}
Where $\phi\left(x\right)$ is the B-L potential, $n_{n}\left(x\right)$
is the neutron density of the object, $n_{0}e^{\left(\pm\right)\phi\left(x\right)/kT_{0}}$
is the density distribution of relic scalar (anti)neutrinos in a B-L
potential, k is Boltzmann constant and $T_{0}$ is temperature of
relic scalar neutrino and anti neutrino background. The net compensated
B-L charge of the macroscopic object is given as:}

\textcolor{black}{
\begin{equation}
b=\underset{V}{\int}\left(n_{n}\left(x\right)+n_{0}\left(x\right)e^{-\phi\left(x\right)/kT_{0}}-n_{0}e^{\phi\left(x\right)/kT_{0}}\right)dV,
\end{equation}
where the integration is over the volume of the object. Since it is
very difficult to find the analytical solution of the differential
equation (3) due to improper values of parameter $\left(n_{n}/n_{0}\sim10^{21}\right)$,
we restrict ourselves to a qualitative analysis at this stage. }

\textcolor{black}{The total energy of the compensating scalar neutrino
in a spherical macroscopic object is negative, so that the potential
energy of this neutrinos is greater than their kinetic energy $(-U)>KE$.
If we assume that scalar neutrinos obey the Boltzmann distribution
at temperature $T_{0}\mathcal{=O}(\thicksim300\lyxmathsym{\textdegree}K)$
for Earth and $T_{0}=\mathcal{O}(\thicksim10^{4}\lyxmathsym{\textdegree}K)$
for Sun, where the temperatures are thermalization temperatures for
the sneutrinos within the earth and the sun, respectively, then their
mean kinetic energy is $<KE>=(3/2)kT_{0}$ and their potential energy
in terms of the compensated B-L charge is given as $U=-\lyxmathsym{\textgreek{a}}_{B-L}(N_{n}-N_{\lyxmathsym{\textgreek{n}}}+N_{\bar{\lyxmathsym{\textgreek{n}}}})/R$,
where R is the object radius, $N_{n}$, $N_{\lyxmathsym{\textgreek{n}}}$
and $N_{\bar{\lyxmathsym{\textgreek{n}}}}$ are the B-L charges due
to neutrons, scalar neutrinos and scalar antineutrinos within the
object. The contribution of $N_{\bar{\lyxmathsym{\textgreek{n}}}}$
is negligible, so that we can write the following equation:}

\textcolor{black}{
\begin{equation}
{\color{red}{\color{black}\lyxmathsym{\textgreek{a}}_{B-L}\frac{N_{n}-N_{\lyxmathsym{\textgreek{n}}}}{R}>\frac{3}{2}kT_{0}}{\color{black}.}}
\end{equation}
}

\textcolor{black}{The neutrino emission from the sun will constantly
decrease the Sun's lepton number. This loss is compensated by the
relic sneutrinos present in the enviroment around the Sun.}

\textcolor{black}{The massless hypothetic photon $\lyxmathsym{\textgreek{g}}_{B-L}$
should produce a sort of effective repulsive \textquotedblleft{}Coulomb\textquotedblright{}
force around the earth. Therefore, the interaction between the earth
and a test object with mass $m_{i}$ and compensated B-L charge $b_{i}$
can be expressed as follows:}

\textcolor{black}{
\begin{equation}
F=-G\frac{Mm_{i}}{r^{2}}+\lyxmathsym{\textgreek{a}}_{B-L}\frac{Bb_{i}}{r^{2}},
\end{equation}
where $B$ and $M$ are the compensated B-L charge and mass of the
earth. If we define the compensation rate as $\lyxmathsym{\textgreek{b}}=(N_{n}-N_{\lyxmathsym{\textgreek{n}}})/N_{n}$,
using eq. (5), $\lyxmathsym{\textgreek{b}}$ becomes}

\textcolor{black}{
\begin{equation}
{\color{red}{\color{black}{\color{red}{\color{black}\lyxmathsym{\textgreek{b}}>6,6\frac{(R/cm)}{\alpha_{B-L}N_{n}}(T_{0}/}}\text{\textdegree}K)}}
\end{equation}
If we write $B$ and $b_{i}$, in terms of $\lyxmathsym{\textgreek{b}}$
and $\lyxmathsym{\textgreek{b}}_{i}$, then the force equation (6)
can be written as }

\textcolor{black}{
\begin{equation}
F=-G\frac{Mm_{i}}{r^{2}}+\lyxmathsym{\textgreek{b}}\lyxmathsym{\textgreek{b}}_{i}\lyxmathsym{\textgreek{a}}_{B-L}\frac{N_{n}N_{n_{i}}}{r^{2}}.
\end{equation}
}

\textcolor{black}{In Table 1 the total neutron number, compensation
rates and effective interaction constants ($\lyxmathsym{\textgreek{a}}^{eff}=\lyxmathsym{\textgreek{b}}\lyxmathsym{\textgreek{b}}_{i}\lyxmathsym{\textgreek{a}}_{B-L}$)
of the earth and the sun \cite{Brown,Sun} are given. In the experiments,
the compensation rates of the test object and earth must be taken
to equal each other. By using the data given in Table 1, lower limit
for the B-L interaction constant can be obtained as }

\textcolor{black}{
\begin{equation}
{\color{red}{\color{black}\lyxmathsym{\textgreek{a}}}_{{\color{black}B-L}}{\color{black}>}{\color{black}\{}{\color{black}\begin{array}{c}
0.27\times10^{-29}Earth\\
{\color{black}{\color{black}{\color{red}{\color{black}{\color{red}{\color{black}{\color{red}{\color{black}{\color{red}{\color{black}0.38}{\color{black}\times10^{{\color{black}-{\color{black}31}}}}{\color{black}Sun}}}}}}}}}}{\color{white}n}
\end{array}}{\color{black}.}}
\end{equation}
}Let us remind that the Eq. (9) holds if the compensation mechanism
at work.

\textcolor{black}{The contribution to the $\bar{\lyxmathsym{\textgreek{n}}}_{e}e$
interaction cross section from the $\lyxmathsym{\textgreek{g}}_{B-L}$
proposed by the model should be less than W and Z boson contribution,
so that the upper limit on the B-L interaction constant may be determined
by $\bar{\lyxmathsym{\textgreek{n}}}_{e}e$ interaction. Whereof it
follows that $\lyxmathsym{\textgreek{a}}_{B-L}<G_{F}s/\sqrt{2\pi}$
with $E_{\lyxmathsym{\textgreek{n}}}~10\mbox{ \ensuremath{MeV}}$
and $s\approx2E_{\lyxmathsym{\textgreek{n}}}m_{e}~10^{-5}$$GeV^{2}$
we have approximately $\lyxmathsym{\textgreek{a}}_{B-L}<10^{-12}$.}

\textcolor{black}{}
\begin{table}
\textcolor{black}{\caption{Some parameters and compensation rates for the earth and the sun}
}%
\begin{tabular}{|c|c|c|}
\hline 
 & \textcolor{black}{Earth} & \textcolor{black}{Sun}\tabularnewline
\hline 
\hline 
\textcolor{black}{$R(cm)$} & \textcolor{black}{$6.4\times10^{8}$} & \textcolor{black}{$7\times10^{10}$}\tabularnewline
\hline 
\textcolor{black}{$N_{n}$} & \textcolor{black}{$1.66\times10^{51}$} & \textcolor{black}{$1.72\times10^{56}$}\tabularnewline
\hline 
\textcolor{black}{$\beta\alpha_{B-L}$} & \textcolor{black}{$7.63\times10^{-40}$} & \textcolor{black}{${\color{black}{\color{black}{\color{black}{\color{black}{\color{red}{\color{black}2}{\color{black}.}{\color{black}68}{\color{black}\times}{\color{black}1}{\color{black}0}^{{\color{black}{\color{black}-}41}}}}}}}$}\tabularnewline
\hline 
\textcolor{black}{$\alpha^{eff}$} & \textcolor{black}{$<2.1\times10^{-49}$} & \textcolor{black}{$<5.3\times10^{-49}$}\tabularnewline
\hline 
\end{tabular}

\end{table}

\textcolor{black}{To form the B-L model it has been added an $U(1)_{B-L}^{'}$
group to the Standard Model (SM) gauge group as $SU(3)_{C}\times SU(2)_{W}\times U(1)_{Y}\times U(1)_{B-L}^{'}.$
Hence the covariant derivative has been as follows:}

\textcolor{black}{
\begin{equation}
\partial_{\lyxmathsym{\textgreek{m}}}\rightarrow D_{\lyxmathsym{\textgreek{m}}}=\partial_{\lyxmathsym{\textgreek{m}}}-ig_{2}\mathbf{T\cdot A_{\lyxmathsym{\textgreek{m}}}}-ig_{1}\frac{Y}{2}B_{\lyxmathsym{\textgreek{m}}}-i{\color{black}{\color{red}{\color{black}{\color{red}{\color{black}g}}}}}_{{\color{black}{\color{red}{\color{black}1}}}}^{{\color{black}{\color{red}{\color{red}{\color{black}'}}}}}bB_{\lyxmathsym{\textgreek{m}}}^{'}.
\end{equation}
}

\textcolor{black}{Where $g_{1},g_{2}$ and $g_{1}^{'}=\sqrt{4\pi\alpha_{B-L}}$
are gauge coupling constants; }\textbf{\textcolor{black}{$\mathbf{T}$}}\textcolor{black}{{}
is an isospin operator of a corresponding multiplet of fermion or
Higgs fields; $Y$ is the weak hypercharge, $b$ is the B-L charge
of the corresponding multiplet, $\mathbf{A_{\lyxmathsym{\textgreek{m}}}},B_{\lyxmathsym{\textgreek{m}}}$
and $B_{\lyxmathsym{\textgreek{m}}}^{'}$ are gauge fields. The B-L
charge is $1/3$ for quarks, $-1$ for leptons \cite{Basso}.}

\textcolor{black}{The interaction Lagrangian can be written as }

\textcolor{black}{
\begin{equation}
L=L_{SM}+g_{1}^{'}J_{fer}^{\lyxmathsym{\textgreek{m}}}B_{\lyxmathsym{\textgreek{m}}}^{'},
\end{equation}
where $L_{SM}$ is the Lagrangian of the Standard Model, with right-handed
neutrino, $\nu_{R}$, added}

\textcolor{black}{
\begin{equation}
J_{fer}^{\lyxmathsym{\textgreek{m}}}=\underset{f}{\sum}b_{f}\bar{f}\lyxmathsym{\textgreek{g}}^{\lyxmathsym{\textgreek{m}}}f
\end{equation}
and $b_{f}$ is the B-L charge of the corresponding fermion. The model
proposed is anomaly free.}

\textcolor{black}{In conclusion, we have shown that the gauging of
B-L charge without compensation admits coupling constant to be $\alpha_{B-L}<10^{-49}$.
If the compensation mechanism takes place, the interaction constant
of the B-L charge may have another interval, namely, ${\color{black}10^{-29}<\lyxmathsym{\textgreek{a}}_{B-L}<10^{-12}}$. }

\end{document}